\documentclass[11pt]{article}
\usepackage{a4wide}
\usepackage{cite}

\usepackage[dvipdfm]{graphicx}
\usepackage{amssymb}

\usepackage{xcolor}

\usepackage{amsmath}
\usepackage{slashed}
\usepackage{hyperref}
\hypersetup{
    unicode=true,
    colorlinks=true,
    linkcolor=blue,
    citecolor=red,
    filecolor=black,
    urlcolor=blue
    }

\begin{document}

\title{
Non-thermal Higgs Spectrum in Reheating Epoch:\\
Primordial Condensate vs.\ Stochastic Fluctuation
}

\author{
Kunio Kaneta\thanks{E-mail: \tt kkaneta@lab.twcu.ac.jp}\ {} and
Kin-ya Oda\thanks{E-mail: \tt odakin@lab.twcu.ac.jp}\bigskip\\
$^{*\dagger}$\it\normalsize Department of Mathematics, Tokyo Woman’s Christian University\\
\it\normalsize Tokyo 167-8585, Japan
}

\maketitle

\begin{abstract}\noindent
Since electroweak symmetry is generally broken during inflation, the Standard Model Higgs field can become supermassive even after the end of inflation.
In this paper, we study the non-thermal phase space distribution of the Higgs field during reheating, focusing in particular on two different contributions: primordial condensate and stochastic fluctuations.
We obtain their analytic formulae, which agree with the previous numerical result.
As a possible consequence of the non-thermal Higgs spectrum, we discuss perturbative Higgs decay during reheating for the case it is kinematically allowed.
We find that the soft-relativistic and hard spectra are dominant in the decay rate of the stochastic fluctuation and that the primordial condensate and stochastic fluctuations decay almost at the same time.
\end{abstract}

\newpage
\section{Introduction}

Despite the discovery of the Higgs boson at the Large Hadron Collider at CERN \cite{ATLAS:2012yve,CMS:2012qbp}, the least understood in the Standard Model of particle physics is the Higgs sector.
This is simply because, in the absence of sufficient experimental data, many possibilities can fit into the theory of electroweak symmetry breaking, which necessarily include cosmological implications; see e.g.\ Ref.~\cite{Lebedev:2021xey} for a recent review.
For instance, electroweak phase transition in the early universe has been extensively studied and many connections to other subjects including baryogenesis \cite{Kuzmin:1985mm,Turok:1990in,Bochkarev:1990fx,Turok:1990zg,Nelson:1991ab,Turok:1991uc,Cohen:1993nk,Funakubo:1995kw,Cline:1995dg,Rubakov:1996vz,Funakubo:1996dw,Trodden:1998ym,Fromme:2006cm,Aoki:2008av,Guo:2016ixx,Enomoto:2021dkl,Enomoto:2022rrl} and gravitational waves \cite{Grojean:2006bp,Caprini:2007xq,Huber:2008hg,Espinosa:2010hh,No:2011fi,Binetruy:2012ze,Kikuta:2014eja,Kakizaki:2015wua,Caprini:2015zlo,Dorsch:2016nrg,Caprini:2018mtu,Ellis:2018mja,Caprini:2019egz,Zhou:2020irf} have been found.
The Higgs sector may also be responsible for producing dark matter, known as the Higgs portal scenario \cite{Silveira:1985rk,McDonald:1993ex,Burgess:2000yq,Davoudiasl:2004be,Mambrini:2011ik,Baek:2011aa,Djouadi:2011aa,Lebedev:2011iq,Djouadi:2012zc,Baek:2012uj,Baek:2012se,Han:2015hda,Casas:2017jjg,Ellis:2017ndg}; see also Ref. \cite{Mambrini2021} for a pedagogical introduction on particle dark matter.

The recent development of observation technologies allows us to address the cosmological history at much earlier times than when the electroweak phase transition takes place.
The inflationary scenario is one of the main targets of the recent cosmological/astrophysical observations; see e.g.\ Refs.~\cite{Olive:1989nu,Linde:1990flp,Lyth:1998xn,Linde:2000kn,Martin:2013tda,Martin:2013nzq,Martin:2015dha} for reviews.
Indeed, cosmic microwave background (CMB) observation has set constraints on the scalar and tensor power spectra, by which many inflationary models have already been ruled out~\cite{Planck:2018jri}.
In such early times during the inflation epoch, the Higgs field may play an important role.
One such example is the Higgs inflation where the Standard Model Higgs is identified as the inflaton \cite{Salopek:1988qh,Bezrukov:2007ep}; see also Refs.~\cite{Bezrukov:2010jz,Hamada:2014iga,Bezrukov:2014bra,Hamada:2014raa,Hamada:2015ria,Hamada:2016onh,Jinno:2017lun,Jinno:2019und,Lee:2020yaj,Rubio:2018ogq} for related topics and a review.

In the case that the Higgs itself is not the inflaton and the inflation is driven by another field, the Higgs is regarded as a light spectator field during inflation/reheating.
In this case, the Higgs vacuum stability during inflation and reheating is an important subject.
The mass measurement of the Higgs and top quark has brought up an indication that the vacuum becomes unstable at high energies, whose cosmological implications have also been extensively studied \cite{Espinosa:2015qea,Herranen:2015ima,Kohri:2016wof,East:2016anr,Enqvist:2016mqj,Joti:2017fwe,Espinosa:2018eve,Markkanen:2018pdo}.
The Higgs field may also affect in many ways the temperature evolution after the end of inflation.
For instance, the highest temperature of the universe is affected by the gravitational production of the Higgs during reheating \cite{Clery:2021bwz,Clery:2022wib} or preheating through thermalization of weak gauge bosons produced by the oscillation of primordial Higgs condensate~\cite{Enqvist:2013kaa,Enqvist:2014bua,Figueroa:2015rqa,Passaglia:2021upk}.

In discussing the cosmological role of the Higgs, especially during and right after inflation, there are two distinctive contributions: the primordial condensate and the stochastic fluctuations.
Sometimes their difference is not well recognized or even mixed up.
This is partly because when one does not need to care about the momentum distribution of the primordial Higgs field, such refined separation is not necessary.
However, when one needs to care, e.g.\ in considering the decay of the Higgs, then a closer look at the difference is needed.

In the present work, we focus on the non-thermal phase space distribution arising from two different origins, namely, the primordial Higgs condensate and the stochastic fluctuation.
These are both seen as a displacement of the Higgs field value during inflation, hence are hardly distinguishable from each other in the field space.
On the other hand, they have different phase space distributions after the end of inflation.
We will explicitly show such a non-thermal spectrum of the Standard Model Higgs and, as a possible implication, will discuss their subsequent decay.

The paper is organized as follows. 
In Sec.~\ref{sec: Higgs and inflation sectors}, we first present an explicit model of the Higgs and inflaton sectors to set up the stage for our discussion.
We explain the dynamics of the Higgs during inflation in Sec.~\ref{sec: Higgs dynamics during inflation}, where we first clarify what the primordial condensate means and then briefly review the stochastic fluctuations.
The non-thermal Higgs spectrum after the end of inflation is discussed in Sec.~\ref{sec: Higgs spectrum after inflation}, which is then applied to the non-thermal Higgs decay in Sec.~\ref{sec: Non-thermal Higgs decay}.
The summary of our findings is given in Sec.~\ref{sec: Summary}.

\section{Higgs and inflation sectors}
\label{sec: Higgs and inflation sectors}

Before discussing the detail of the Higgs dynamics during inflation, we make our framework explicit in this section, including the Higgs and inflaton sectors.
While we will consider the Higgs sector in its conventional form (the quadratic and quartic terms in the potential), the choice of the inflationary sector is arbitrary to some extent.
Since what we will argue in the subsequent sections does not quite depend on the inflationary models, our choice here should be considered as an example for the sake of concreteness.

\subsection{Set-up}

The relevant part of the action is given by
\begin{align}
   S &= \int d^4x \sqrt{-g} \left({\cal L}_{\rm grav} + {\cal L}_{\rm inf} + {\cal L}_{\rm Higgs}\right),
\end{align}
where ${\cal L}_{\rm grav}$, ${\cal L}_{\rm inf}$, and ${\cal L}_{\rm Higgs}$ are the Lagrangian densities for the Einstein gravity, inflaton, and Higgs sectors, respectively.
With the reduced Planck mass $M_\text{P}\simeq 2.4\times10^{18}$ GeV, the gravity Lagrangian is given by
\begin{align}
    {\cal L}_{\rm grav} &= -\frac{M_\text{P}^2}{2}R,
\end{align}
where $R$ is the Ricci scalar.
We do not specify the inflaton sector ${\cal L}_{\rm inf}$ at the moment, but the slow-roll inflation is assumed.
The Higgs sector is given by
\begin{align}
    {\cal L}_{\rm Higgs} &= g^{\mu\nu}\left(D_\mu\Phi\right)^\dagger \left(D_\nu \Phi\right) - V(\Phi),
\end{align}
where $\Phi$ denotes the Higgs doublet and
\begin{align}
    V(\Phi) &= -\mu^2 |\Phi|^2 + \lambda\left(\Phi^\dagger\Phi\right)^2 - \xi \left|\Phi\right|^2 R.
\end{align}
We take the Friedmann-Lema\^itre-Robertson-Walker metric  $g_{\mu\nu}={\rm diag}\!\left(1,-a^2,-a^2,-a^2\right)$ with the scale factor $a$.
The Higgs non-minimal coupling to $R$ is also included in the potential.
Note that such a non-minimal coupling affects the kinetic term of the Higgs when we turn to the Einstein frame.
However, as we will see, only the case $\xi \left|\Phi\right|^2\ll M_\text{P}^2$ becomes relevant, and thus we may effectively regard the given $\Phi$ as defined in the Einstein frame and the non-minimal coupling may be perturbatively treated as an effective mass to the Higgs field (especially during inflation).

To illustrate how $\Phi$ evolves during inflation, we simplify the setup by replacing $\Phi\to \chi/\sqrt{2}$ with $\chi$ a real scalar.
The extension to the doublet is straightforward and will be seen later.
The equation of motion for $\chi$ during inflation is given by
\begin{align}
    \ddot{\chi}+3H\dot\chi-a^{-2}\vec\nabla^2\chi + \partial_\chi V(\chi) = 0,
    \label{eq: EoM for chi}
\end{align}
where the dot and $\partial_\chi$ denote the derivatives with respect to the comoving time $t$ and the field $\chi$, respectively, and $\vec\nabla^2=\sum_{i=1}^3\partial_i^2$ is the Laplacian with respect to the comoving coordinates.
The scalar potential in the present case is given by
\begin{align}
    V(\chi) &= -\frac{\mu^2}{2}\chi^2 + \frac{\lambda}{4}\chi^4 - \frac{\xi}{2}\chi^2 R.
\end{align}
Unlike the free scalar case, the quartic coupling induces an effective mass, which we will take a closer look at based on the stochastic approach.

\subsection{Inflationary model}

Before getting into the detail about the Higgs dynamics during inflation, we clarify the setup of the inflaton sector.
Although our analysis does not quite depend on the choice of the inflation model, for the sake of definiteness, we consider the T-model \cite{Kallosh:2013hoa} whose inflaton potential is given by
\begin{align}
    V(\phi) = 6\lambda_\phi M_\text{P}^4\tanh^2\left(\frac{\phi}{\sqrt{6}M_\text{P}}\right),
\end{align}
where the potential coupling $\lambda_\phi$ is determined by observation.
The amplitude of the curvature power spectrum, $A_\text{S}$, determines the inflationary scale as
\begin{align}
    \lambda_\phi &\simeq \frac{18\pi^2A_\text{S}}{6N^2_{\rm CMB}},
\end{align}
where $N=\ln(a_\text{e}/a)$ is the number of e-folds measured from the end of inflation at $a=a_\text{e}$, and $N_{\rm CMB}$ is the time $a=a_{\rm CMB}$ when the CMB scale exits the horizon.
In the following, we use $\ln(10^9 A_\text{S})=3.044$ measured at the pivot scale $k_{\rm CMB}=0.05$ Mpc$^{-1}$ and assume $N_{\rm CMB}=55$, so $\lambda = 2.1\times 10^{-11}$.
Note that the inflaton starts oscillating about $\phi=0$ after the end of inflation, during which the inflaton potential is approximately quadratic:
\begin{align}
    V(\phi\ll\phi) &\simeq \frac{1}{2}m_\phi^2\phi^2,
\end{align}
where $m_\phi=\sqrt{2\lambda_\phi}M_\text{P}\simeq 1.5\times10^{13}$ GeV.

\section{Higgs dynamics during inflation}
\label{sec: Higgs dynamics during inflation}

Our focus is on the evolution of the Standard Model Higgs field during inflation, where the Higgs is considered as a spectator field whose energy density is assumed to be subdominant compared with that of the inflaton.

\subsection{Primordial condensate}

We first consider the dynamics of the what-we-call \emph{primordial condensate}, a.k.a.\ misalignment contribution, where the initial field value of $\chi$ is set at some time during or before inflation.
In particular, we define the primordial condensate in a narrow sense by considering that the contribution arises only from the zero momentum state.

Note that we may also consider non-zero amplitudes (and initial phases depending on the wave number) for higher momentum states, i.e., a finite initial number of the corresponding particle.
On the other hand, as we will see shortly, the higher momentum states may acquire a non-zero variance through stochastic fluctuations even when the initial field value (and/or phase) is zero, i.e., when there are no initial particles, or equivalently when we start from the Bunch-Davies vacuum~\cite{Chernikov:1968zm,Bunch:1978yq}.
So, the initial displacement would add up to the stochastic fluctuation in the end, if it may be regarded as a small correction to the Bunch-Davies vacuum.
Notice however that if the initial amplitude is much greater than the stochastic fluctuation, the Fokker-Planck equation for a Gaussian noise may not be a good approximation any further, since the noise term may be affected by the primordial contribution, and hence deviates from Gaussian distribution,\footnote{
In such a case, the initial condition may not be close to the Bunch-Davies vacuum. See e.g.\ Ref.~\cite{Kundu:2011sg} for deviations from the Bunch-Davies vacuum.
}
which is beyond the scope of our present work.
Therefore in the following argument, we constrain ourselves to the case where only the zero momentum state may form the primordial condensate.\footnote{
This assumption is justified when e-folding from an ``initial'' time $a_{\rm ini}$ to $a_{\rm CMB}$ is sufficiently large, $\ln(a_{\rm CMB}/a_{\rm ini})\gtrsim\ln(M_\text{P}/H_{\rm inf})$ barring the existence of primordial condensates that have wave number larger than $M_\text{P}$. Without such trans-Plackian modes, all the primordial condensates are red-shifted to sufficiently low momenta that can be collectively approximated by the zero momentum mode as in our treatment.
}

For the zero-momentum mode $\chi$, the equation of motion is given by
\begin{align}
    \ddot\chi + 3H\dot\chi + \partial_{\chi} V(\chi) = 0.
    \label{eq: EoM for chi0}
\end{align}
Here, we approximate the Hubble parameter during inflation as a constant such that $H_{\rm inf}$ is the same as the one at the end of inflation $a=a_\text{e}$ i.e.\ $H_{\rm inf}=H(a=a_\text{e})\equiv H_\text{e}$.
Supposing that $\chi$ has a vanishing velocity $\dot\chi=0$ at an initial time, there are two cases for the solution of Eq.~(\ref{eq: EoM for chi0}).
If the potential term is negligibly small compared with the Hubble friction term, then $\chi(t)=\chi_0$, where $\chi_0$ is a constant field value set at an initial time, so the Higgs field value remains constant during inflation.
On the other hand, if the potential term dominates over the Hubble friction term, then
\begin{align}
\chi(t)\simeq \chi_0\ {\rm sn}\!\left(K\!\left(-1\right)-\sqrt{\lambda/2}\,\chi_0t,\ -1\right),
\end{align}
where ${\rm sn}(u,m)$ is the elliptic sine and $K\!\left(n\right)$ is the complete elliptic integral of the first kind: $K\!\left(-1\right)\simeq 1.31$. We see that $\chi(t)$ oscillates during inflation while the Hubble friction still damp the amplitude.
Therefore, at some time when the Hubble drag becomes efficient, $\chi(t)$ settles down to a certain constant value.

\begin{figure}
    \centering
    \includegraphics[width=0.8\textwidth]{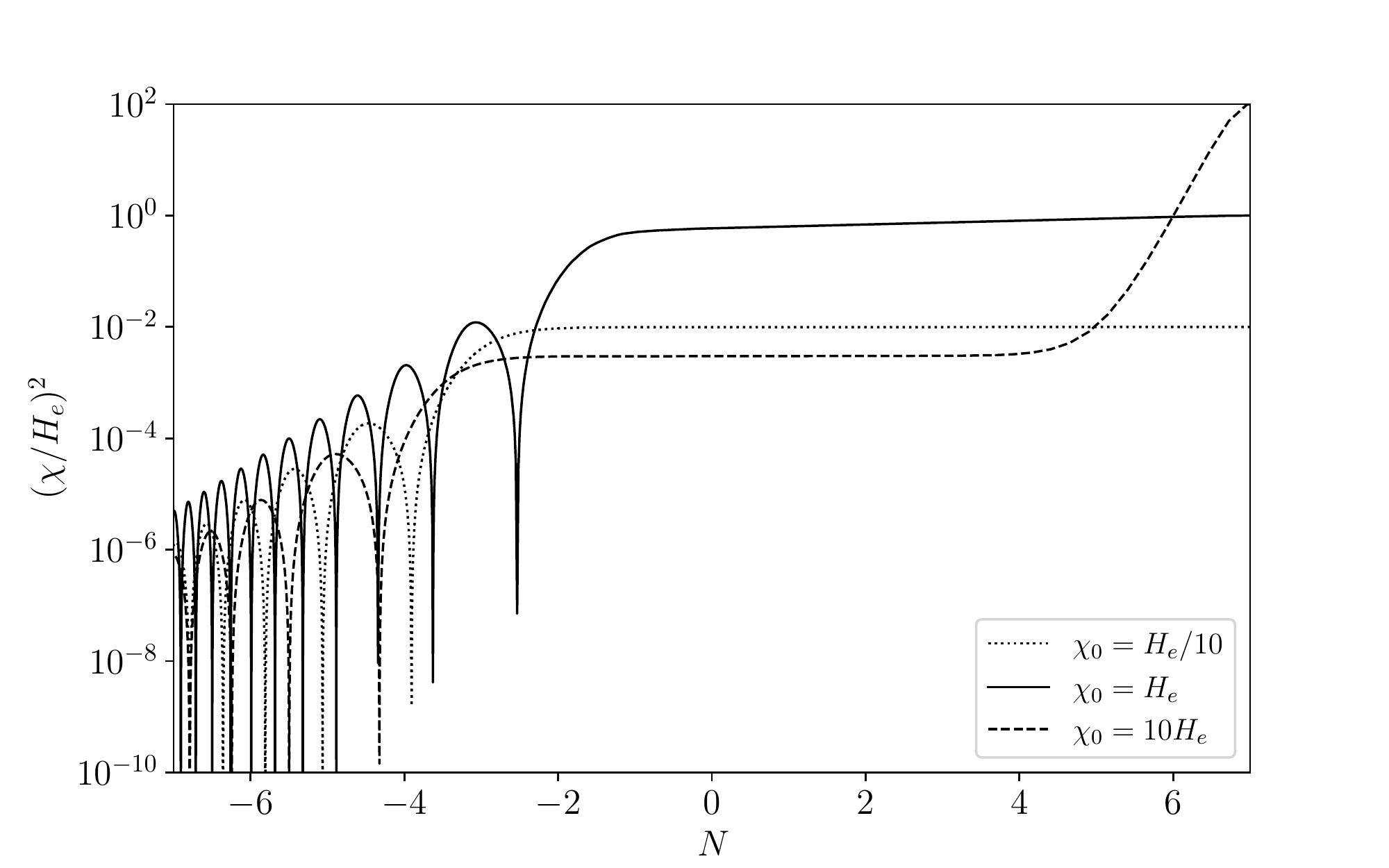}
    \caption{Evolution of the primordial condensate for three different initial amplitudes as a function of $N=\ln(a_\text{e}/a)$. The squared amplitude is normalized by $H_\text{e}=m_\phi$. The time flows from right to left. 
    The case $\chi_0=10H_\text{e}$ shows the initial settle down to the constant value, whereas the other two cases correspond to the dominance of the Hubble friction term from the beginning; see main text.
    }
    \label{fig: misalignment}
\end{figure}

Figure~\ref{fig: misalignment} shows the evolution of the primordial condensate for the different choices of the initial amplitude.
It shows that a non-zero initial amplitude ends up with a constant final value during inflation if the inflation lasts long enough, as expected.
After the end of inflation, $\chi(t)$ starts oscillating when the Hubble friction drops below the potential term.

\subsection{Stochastic fluctuations}
In addition to the primordial condensate, the stochastic fluctuations, which always exist even if the initial amplitude of the Higgs is zero, also affect the Higgs field value during inflation.
We briefly review how the Higgs field acquires an effective vacuum expectation value and hence mass during inflation, based on the stochastic approach~\cite{Starobinsky1986,Starobinsky1994}.\footnote{
See e.g.\ Refs.~\cite{Fujita:2017lfu,Pinol:2018euk,Pinol:2020cdp,Cable2021,Fujita:2022fit,Honda:2023unh} for related topics.
}

The stochastic approach begins by dividing the field into two pieces, one of which resides in the infrared (IR) domain, and the other in the ultraviolet (UV) domain.
This can be done in practice by introducing a cutoff scale~$k_{\rm cut}$ in the comoving momentum space.\footnote{
Note that physical momentum $p$ is related to comoving momentum $k$ through $p=k/a$.
}
It is convenient to discuss with the mode function $\chi_{\vec k}$ (or $u_k$) defined by 
\begin{align}
    \chi(t,\vec x) &= \int \frac{d^3k}{(2\pi)^{3/2}}e^{i\vec k\cdot \vec x}\chi_{\vec k}(t),
\end{align}
where
\begin{align}
    \chi_{\vec k}(t) &= a_{\vec k} u_k(t) + a^\dagger_{-\vec k}u^*_k(t),
\end{align}
in which $a_{\vec k}$ and $a^\dagger_{-\vec k}$ are the annihilation and creation operators.
In the same manner, we define the conjugate momentum by $\pi \equiv \dot\chi$ whose mode function is $\pi_{\vec k}(t) = a_{\vec k}v_k(t)+a^\dagger_{\vec k}v^*_k(t)$.
The splitting can be realized by introducing a window function $W_k(t)$ such that
\begin{align}
    u_k(t) = u_{k<k_{\rm cut}}(t) + u_{k>k_{\rm cut}}(t),
\end{align}
where
\begin{align}
    u_{k>k_{\rm cut}}
        &\equiv \left[1- W_k(t)\right]u_k,\\
    u_{k<k_{\rm cut}} 
        &\equiv W_k(t)u_k.
    \label{eq: u_UV and u_IR}
\end{align}
A conventional choice for $W_k(t)$ is given by the step function as
\begin{align}
    W_k(t) &= \theta\!\left(\epsilon a(t) H - k\right),
\end{align}
where $\epsilon\ll1$ is an arbitrary constant, corresponding to $k_{\rm cut}=\epsilon a H$.

Denoting the IR contributions of $\chi$ and $\pi$ as
\begin{align}
    \overline\chi &\equiv
    \int \frac{d^3k}{(2\pi)^{3/2}}e^{i\vec k\cdot\vec x}W_k\chi_{\vec k},\\
    \overline{\pi} &\equiv 
    \int \frac{d^3k}{(2\pi)^{3/2}}e^{i\vec k\cdot\vec x}W_k\pi_{\vec k},
\end{align}
we see that
\begin{align}
    \dot{\overline\chi} - \overline\pi = f^\chi,
    \label{eq: dchi/dt - pi}
\end{align}
which is non-zero, where
\begin{align}
    f^\chi(t,\vec x) &\equiv \int \frac{d^3k}{(2\pi)^{3/2}}e^{i\vec k\cdot\vec x} \dot W_k \chi_k.
    \label{eq: noise for chi}
\end{align}
When $H^2\gg \partial_\chi^2 V$, the IR modes ($k<k_{\rm cut}$) are in the slow-roll regime,\footnote{
See e.g.\ Ref.~\cite{Chiba:2009sj}.
}
and thus, we obtain the relation among the IR-only quantities:
\begin{align}
    \overline \pi &\simeq -\frac{\partial_{\overline \chi} V(\overline\chi)}{3H}.
    \label{eq: slow-roll for pi}
\end{align}
Plugging Eq. (\ref{eq: slow-roll for pi}) into Eq. (\ref{eq: dchi/dt - pi}), we end up with
\begin{align}
    \dot{\overline{\chi}} &\simeq -\frac{\partial_{\overline\chi}V(\overline\chi)}{3H} + f^\chi.
    \label{eq: Langevin for chi}
\end{align}
The appearance of non-zero $f^\chi$ indicates that UV modes keep flowing into the IR domain during inflation, caused by the constant (Hubble) horizon size of the (quasi) de Sitter spacetime.

It is important that Eq.~(\ref{eq: Langevin for chi}) may be regarded as the Langevin equation for $\overline\chi$ with a random noise $f^\chi$.
Furthermore, the distribution of the noise is Gaussian as the variance is given by
$
\left\langle  f^\chi(t_1,\vec x)f^\chi(t_2,\vec x)\right\rangle = \left(H_\text{e}^3/4\pi^2\right)\delta(t_1-t_2)
$
when $\chi_k$ in Eq.~(\ref{eq: noise for chi}) follows the equation of motion for a free massless state, whose mode functions are given by
$
u_k=-(H/\sqrt{2k})\left(\eta-i/k\right)e^{-ik\eta}
$
for the conformal time $\eta=-1/aH$.\footnote{
For reader's ease, mass dimensions are $[a]=0$, $[\eta]=-1$, $[u_k]=-1/2$, $[a_{\vec k}]=-3/2$, $[W_k]=0$, $[\dot W_k]=1$, $[\chi_k]=-2$, $[\chi]=1$, $[\overline\chi]=1$, and $[f^\chi]=2$.
}
Therefore the distribution of $\overline\chi$ in the field space, $P(\overline\chi,t)$, follows the Fokker-Planck equation given by\footnote{
See e.g.\ Chapter~4 in Ref.~\cite{Zinn-Justin:1989rgp}.
}
\begin{align}
    \dot P(\overline\chi,t) &= \frac{1}{2}\partial_{\overline\chi}\left(
        \frac{H_\text{e}^3}{8\pi^2}\partial_{\overline\chi}+\frac{\partial_{\overline\chi}V(\overline\chi)}{3H_\text{e}}
    \right)P(\overline\chi,t).
    \label{eq: Fokker-Planck for chi}
\end{align}

Supposing that the duration of the inflation is long enough so that $\overline\chi$ reaches the equilibrium state in the field space,\footnote{
This is indeed the case if the number of e-folds for the duration of the inflation satisfies $N\gtrsim 1/\sqrt{\lambda}$ \cite{Hardwick:2017fjo}.
} 
we can readily find the solution of Eq.~(\ref{eq: Fokker-Planck for chi}) as
\begin{align}
    P(\overline\chi) &= N \exp\left[
        -\frac{8\pi^2 V(\overline\chi)}{3H_\text{e}^4}
    \right],
\end{align}
where the normalization factor $N$ is determined by $\int d\overline\chi P(\overline\chi)=1$.
For instance, when the potential energy is dominated by the quartic coupling, $V(\overline\chi)\simeq (\lambda/4)\overline\chi^4$, we find
\begin{align}
    \langle \overline\chi^2\rangle &=
    \frac{\sqrt{3}}{\Gamma^2(\frac{1}{4})}\frac{H_\text{e}^2}{\sqrt{\lambda}}\simeq 0.132\times \frac{H_\text{e}^2}{\sqrt{\lambda}}.
\end{align}
Through the non-zero variance of the IR modes, $\chi$ acquires an effective mass for the separation $\chi^2=\langle\overline\chi^2\rangle+\delta\chi^2$ from the quartic coupling, namely, $m_{\chi,{\rm eff}}^2 \simeq \lambda \langle\overline\chi^2\rangle$.

The Standard Model Higgs may obtain the non-zero variance in the same manner, except that $\Phi$ has four real scalar degrees of freedom.
Therefore, we find
\begin{align}
    \langle \overline\Phi^\dagger\overline\Phi \rangle &=
    \sqrt{\frac{3}{8\pi}}\frac{H_\text{e}^2}{\sqrt{\lambda}}\left(
        1 - (\pi-4)\sqrt{\frac{2\pi}{3}}\frac{\mu_H^2}{H_\text{e}^2} + \dots
    \right),
\end{align}
where $\overline\Phi$ denotes the IR modes of $\Phi$, and $\mu_H^2\equiv -\mu^2 + 6\xi H_\text{e}^2$ with $R=-12H_\text{e}^2$, which we have assumed $|\mu_H^2|\ll H_\text{e}^2$.
When the Higgs potential is dominated by the quartic coupling during inflation, the effective mass $m_{h,{\rm eff}}$ is given by
\begin{align}
    m_{h,{\rm eff}}^2 = \sqrt{\frac{3\lambda }{8\pi}} H_\text{e}^2
\end{align}
for every single real scalar in $\Phi$.
In the same manner, the total energy density becomes
\begin{align}
    \rho_{h} &\simeq \left\langle V(\overline\Phi)\right\rangle = \frac{3}{8\pi^2}H_\text{e}^4.
    \label{eq: rho_h(a_e) for the stochastic fluctuation}
\end{align}
Finally, it should be stressed that the stochastic fluctuation is induced by the gravitational effect and thus always exists even in the absence of the primordial Higgs condensate.

\section{Higgs spectrum after inflation}
\label{sec: Higgs spectrum after inflation}

We turn our discussion to the phase space distribution of the Higgs after the end of inflation.
In the following argument, we suppose that the thermalization of the Higgs takes a very long time such that it does not complete and that the non-thermal spectrum persists.
This setup is justified e.g.\ when the reheating temperature is sufficiently low compared with the effective Higgs mass, which can easily be realized unless $\lambda$ is extremely small.

The phase space distribution for the primordial condensate is particularly simple, which is by definition given by
\begin{align}
    f_{\rm cond}(p,t) &= n_{\rm cond} \left(2\pi\right)^3\delta^3(\vec p),
    \label{f for primordial condensate}
\end{align}
where $p$ is the physical momentum, and the normalization $n_{\rm cond}$ is determined by
\begin{align}
    n_{\rm cond} &=
    \int \frac{d^3p}{(2\pi)^3}f_{\rm cond}(p,t) = \frac{\rho_{h,{\rm cond}}(t)}{m_{h,{\rm eff}}}.
    \label{eq: n_mis}
\end{align}
By denoting the initial squared amplitude of the Higgs by $|\Phi_0|^2$, the effective mass is given by $m_{h,{\rm eff}}^2 = \lambda |\Phi_0|^2$, and the initial energy density $\rho_{h,0}$ is determined accordingly.
Note that $\rho_{h,{\rm cond}}$ keeps to be $\rho_{h,{\rm cond}}\sim \rho_{h,0}$ until $a_{\rm osc}$ that satisfies $H(a=a_{\rm osc})\simeq m_{h,{\rm eff}}$. Afterward, $\rho_{h,{\rm cond}} \simeq \rho_{h,0}(a/a_{\rm osc})^{-4}$ assuming that the quartic term in the potential dominates during the Higgs oscillation.

It becomes more involved to take into account other Higgs spectra than the primordial condensate.
As we will discuss in the following, there are three distinctive regimes in the phase space density for the stochastic fluctuation.

\subsection{Hard spectrum from inflaton scattering}

When $m_{h,{\rm eff}} < m_\phi$, the Higgs quanta may be produced through the inflaton oscillation after the end of inflation.
This process can produce only the Higgs with $k>m_\phi$ for the kinematical reason \cite{Kaneta2022}.
In terms of physical momentum, we have
\begin{align}
    f_{\rm hard}(p,t) &\simeq
    \frac{9\pi}{64}\left(
        \frac{H(t)}{m_\phi}
    \right)^3\left(
        \frac{m_\phi}{p}
    \right)^{9/2}
    \theta\!\left(m_\phi - p\right)\theta\!\left(p - m_\phi a_\text{e}/a\right),
    \label{eq: f_hard}
\end{align}
where we have used $H(t) = H_\text{e} (a/a_\text{e})^{-3/2}$; see also Ref.~\cite{Garcia2021}.
Note that the first step function ensures the energy conservation in the process of $\phi\phi\to\chi\chi$, and the second bounds the spectrum from below by imposing the maximal amount of the redshift of the physical momentum.

\subsection{Soft spectra, relativistic and non-relativistic}

The Higgs spectrum with $k<m_\phi$ is produced via the phase transition from de Sitter (inflation) to matter domination (inflaton oscillation).
The soft spectrum depends on $m_{h,{\rm eff}}$ though the behavior of the effective frequency which is, during reheating, given by 
\begin{align}
   \omega_k^2 &= k^2 + \frac{H_\text{e}^4 m_{h,{\rm eff}}^2}{16}(\eta-\overline\eta_\text{e})^4 - \frac{2}{(\eta-\overline\eta_\text{e})^2},
\end{align}
where $\overline \eta_\text{e} = \eta_\text{e}-2/a_\text{e} H_\text{e}$ \cite{Kaneta2022}.
Writing $(\eta-\overline\eta_\text{e})^2\equiv y$ and $H_\text{e}^4 m_{h,{\rm eff}}^2/16\equiv M^6/3$, we solve
\begin{align}
   k^2 + \frac{M^6}{3}y^2-\frac{2}{y}=0
\end{align}
for $y$, yielding
\begin{align}
   y = \frac{-\left(k/M\right)^2+\left[3+\sqrt{9+\left(k/M\right)^6}\right]^{2/3}}{M^2\left[3+\sqrt{9+\left(k/M\right)^6}\right]^{1/3}} \equiv y_{\rm entry}.
\end{align}
Notice that a given mode $k$ enters the horizon at $y=y_{\rm entry}$.
For $k\gg M$, the horizon entry happens at $y_{\rm entry} \sim k^{-2}\Rightarrow\eta_{\rm entry}\sim 1/k$, which is approximately the massless case.
On the other hand, by keep decreasing $k$ we observe that at some point, $y_{\rm entry}$ becomes independent from $k$, indicating that the spectrum becomes non-relativistic and proportional to $k^{-3}$.
The transition from relativistic to non-relativistic occurs at $k=k_*$ which satisfies
\begin{align}
    y_{\rm entry} (k_*\gg M) \simeq \frac{2}{k_*^2} - \frac{8}{3}\frac{M^6}{k_*^8}
        &= 0&
    &\Rightarrow&
    k_* &= \left(\frac{4}{3}\right)^{1/6}M = \left(\frac{H_\text{e}^2m_{h,{\rm eff}}}{2}\right)^{1/3}. 
    \label{eq: k_*}
\end{align}
The soft spectrum has two regimes: relativistic ($k>k_*$) and non-relativistic ($k<k_*$).

For the relativistic soft spectrum, we may neglect the Higgs mass effect, and thus the phase space distribution is approximately given by the massless case:
\begin{align}
    f_{\rm soft,R}(p,t) &\simeq
    \frac{9}{64}\left(
        \frac{H(t)}{p}
    \right)^6\left(
        \frac{a}{a_\text{e}}
    \right)^3
    \theta\!\left(m_\phi a_\text{e}/a-p\right)\theta\!\left(p-k_*a_\text{e}/a\right).
    \label{eq: f_soft,R}
\end{align}
For $k<k_*$, we obtain
\begin{align}
    f_{\rm soft, NR}(p,t) &\simeq 
    \frac{9}{32}\left(\frac{H}{m_{h,{\rm eff}}}\right)\left(\frac{H}{p}\right)^3\left(\frac{a}{a_\text{e}}\right)^{-3}
    \theta\!\left(k_* a_\text{e}/a-p\right).
    \label{eq: soft,NR}
\end{align}
Therefore the total phase space distribution is given by the sum of Eqs.~(\ref{eq: f_hard}), (\ref{eq: f_soft,R}), and (\ref{eq: soft,NR}):
\begin{align}
    f(p,t) &= f_{\rm hard}(p,t) +  f_{\rm soft,R}(p,t) + f_{\rm soft,NR}(p,t).
    \label{eq: f_total}
\end{align}

\begin{figure}
    \centering
    \includegraphics[width=.8\textwidth]{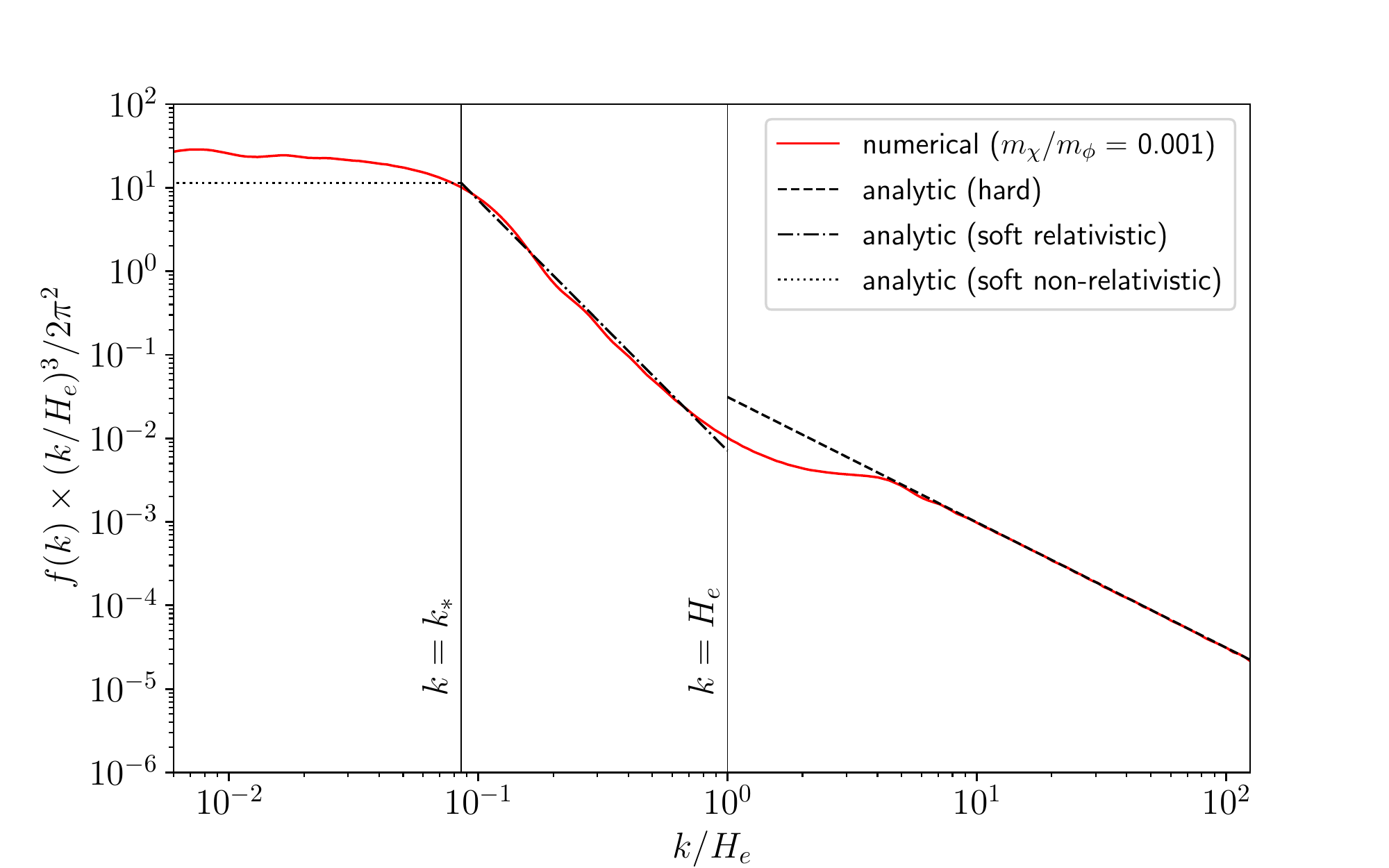}
    \caption{Comparison of non-thermal Higgs spectrum between numerical result~\cite{Kaneta2022} and the analytic estimate.}
    \label{fig: spectrum}
\end{figure}

Figure~\ref{fig: spectrum} shows the total phase space distribution for $\chi$ as a function of the comoving momentum $k=ap$.
The red solid line in the figure is the numerical result taken from Ref. \cite{Kaneta2022}, and the black dashed, dotted, dot-dashed lines are analytical estimates given by Eq.~(\ref{eq: f_total}).
Note that to fit the numerical result by the analytic estimate, we have taken $H_\text{e}=m_\phi/1.25$ which reflects the uncertainty that the end of inflation is not abrupt at $H=H_\text{e}$, rather it is smooth with time.
The choice of $m_\chi/m_\phi=0.001$ corresponds to $\lambda\simeq 10^{-11}$, though our results do not depend on the specific value of $\lambda$ as long as $m_{h,{\rm eff}}<m_\phi$ is satisfied.

\section{Non-thermal Higgs decay}
\label{sec: Non-thermal Higgs decay}

As an implication of the non-thermal Higgs spectrum, we consider the perturbative decay of the Higgs into the Standard Model particles.
This can be possible whenever $m_{h,{\rm eff}}$ is larger than the mass of the decay products.
In such kinematic reasoning, one should keep in mind that the decay products may also acquire their mass from the effective Higgs vacuum expectation value.
Therefore, for instance, the (perturbative) decay into the weak gauge bosons does not occur when $\sqrt{\lambda}\lesssim g$, where $g$ is the SU(2) gauge coupling.
This is normally the case if we extrapolate the renormalization group running of the couplings up to the inflationary scale within the Standard Model \cite{Enqvist:2013kaa,Enqvist:2014bua}.\footnote{
See also, e.g., Refs. \cite{Figueroa:2015rqa,Passaglia:2021upk} for implications of non-perturbative Higgs decay. The Higgs energy density, however, cannot dissipate at all through the parametric resonance, since the instability terminates when the energy density of the produced particles reaches the Higgs energy density. Thus the complete dissipation of the energy density needs a perturbative decay \cite{Kofman:1997yn}.
}
However, such running is affected by, for instance, the error in the top quark mass measurements and/or the appearance of new physics at certain high scales~\cite{Hamada:2017sga}.
Therefore, we take rather an agnostic approach to the choice of $\lambda$, regarding it as a free parameter at high scales.

In the same spirit, we take the Higgs decay width $\Gamma_h$ as a free parameter and will keep track of how the Higgs decays.
To do this, instead of looking at the decay on a mode-by-mode basis, we consider the integrated Boltzmann equation given by
\begin{align}
    \dot n_h + 3Hn_h = R_{h\to AB},
    \label{eq: Boltzmann for n_h}
\end{align}
where the decay rate $R_{h\to AB}$ is given by
\begin{align}
    R_{h\to AB} &= 
    -\int\frac{d^3 p_h}{\left(2\pi\right)^32p_h^0}\frac{d^3 p_A}{\left(2\pi\right)^32p_A^0}\frac{d^3 p_B}{\left(2\pi\right)^32p_B^0}
    \left|{\cal M}_{h\to AB}\right|^2f_h\!\left(p_h,t\right)\left(2\pi\right)^4\delta^4(p_h-p_A-p_B)\nonumber\\
    &= -\gamma \Gamma_h,
\end{align}
in which
\begin{align}
    \gamma &\equiv
    \int \frac{d^3p}{\left(2\pi\right)^3}\frac{m_{h,{\rm eff}}}{\sqrt{m_{h,{\rm eff}}^2 + p^2}}f_h(p,t),
\end{align}
with $f_h(p,t)=4f(p,t)$ by taking four degrees of freedom in $\Phi$ into account, or $f_h(p,t) = f_{\rm cond}(p,t)$ for the primordial condensate.\footnote{
All the four modes acquire the effective mass since it comes from the quartic coupling.
}

In the case of the primordial condensate~\eqref{f for primordial condensate}, the Higgs decay is particularly easy.
By using $\rho_h=m_{h,{\rm eff}}n_h$ from Eq.~(\ref{eq: n_mis}), we may rewrite the Boltzmann equation as
\begin{align}
    \dot \rho_h + 4H\rho_h = -\Gamma_h \rho_h
    \label{eq: Boltzmann for rho_h}
\end{align}
after the onset of the Higgs oscillation, approximating that the quartic coupling dominates in the potential during the oscillation~\cite{Garcia:2020wiy}.
The solution of Eq.~(\ref{eq: Boltzmann for rho_h}) is given by
\begin{align}
    \rho_h(t) &= \rho_{h,0}\left(\frac{a}{a_{\rm osc}}\right)^{-4}e^{-\Gamma_h t},
\end{align}
and thus the decay time can readily be estimated as\footnote{
Here, we have neglected the time dependence in $\Gamma_h$, though the effective Higgs mass may potentially be time-dependent during the Higgs oscillation epoch. To see this effect, a more careful study of every possible decay channel is needed, which is left for future work. For a related study, see Refs.~\cite{Garcia:2020wiy,Garcia:2020eof}.
}
\begin{align}
    t_{\rm dec} &= \Gamma_h^{-1} &
    &\Rightarrow&
    \frac{a_{\rm dec}}{a_\text{e}} &\simeq \left(\frac{3}{2}\frac{H_\text{e}}{\Gamma_h}\right)^{2/3},
    \label{eq: a_dec for misalignment}
\end{align}
where we have used $t=2/3H\simeq\left(2/3H_\text{e}\right)\left(a/a_\text{e}\right)^{3/2}$.

\begin{figure}[t]
    \centering
    \includegraphics[width=.8\textwidth]{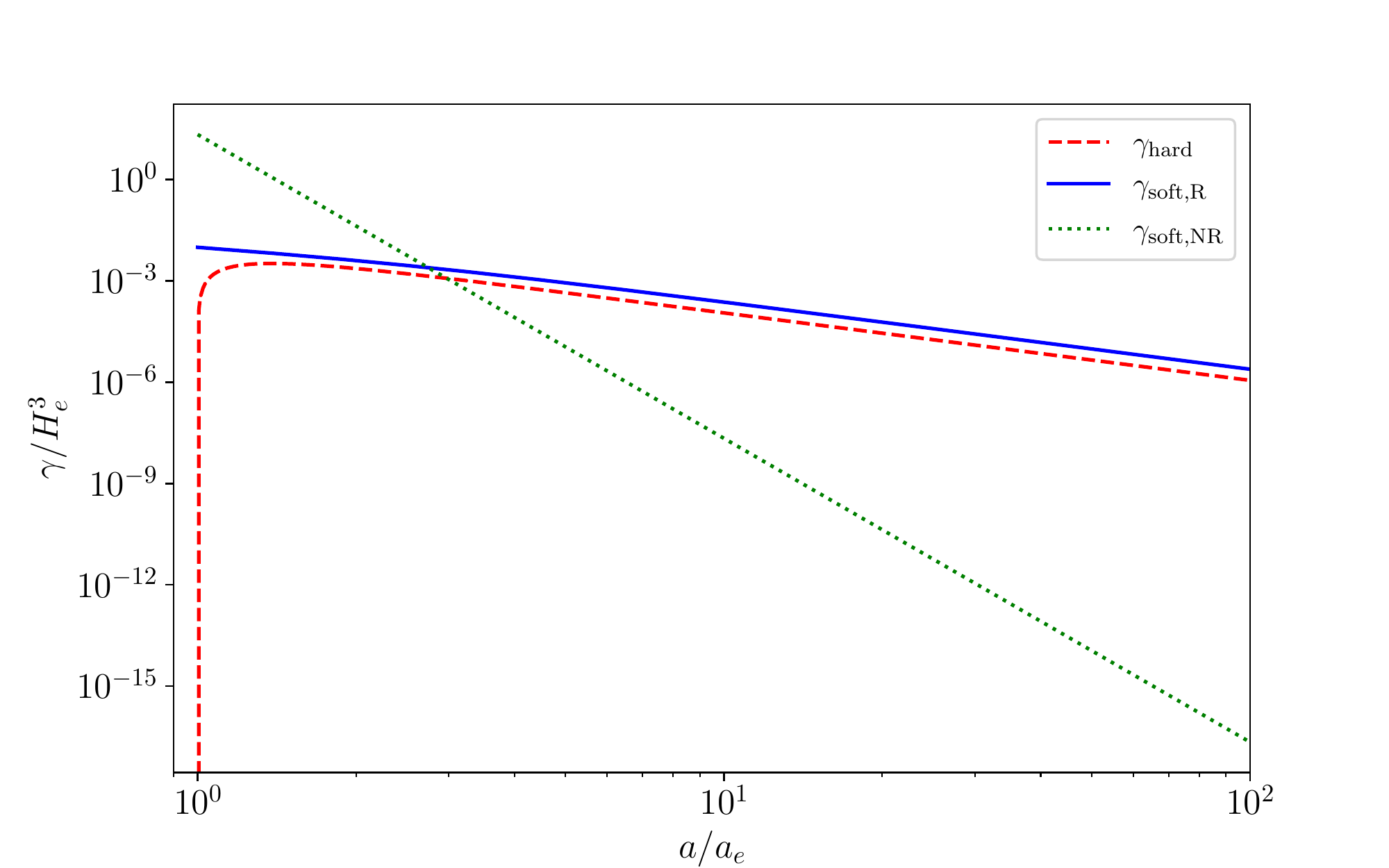}
    \caption{Normalized decay rate of each spectrum (hard, soft relativistic, soft non-relativistic) as a function of the scale factor $a$, with a choice of parameters $\lambda=0.01$ and $H_\text{e}=m_\phi$.}
    \label{fig: gamma}
\end{figure}

The decay of the stochastic fluctuation is more involved, compared with the primordial condensate.
Figure~\ref{fig: gamma} shows each contribution of the spectrum in the decay rate (normalized by $\Gamma_h H_\text{e}^3$) as a function of the scale factor $a$ with $\lambda=0.01$ and $H_\text{e}=m_\phi$.
As seen in the figure, although $\gamma_{\rm soft,NR}$ is the greatest contribution at early times, the other two soon dominate and can be the main contribution in the Higgs decay.
Depending on the value of $\lambda$, they can be comparable, and in the large-$a$ limit we find
\begin{align}
    \gamma_{\rm hard} &\simeq
    \frac{3H_\text{e}^3}{32\pi}\left(
        \frac{a}{a_\text{e}}
    \right)^{-3},\\
    \gamma_{\rm soft,R} &\simeq
    \frac{3}{32\pi^2}\frac{H_\text{e}^4}{m_{h,{\rm eff}}}\left(
        1 - \frac{m_{h,{\rm eff}}H_\text{e}^2}{2m_\phi^3}
    \right)\left(
        \frac{a}{a_\text{e}}
    \right)^{-3},
\end{align} 
where we have used Eq.~(\ref{eq: k_*}) for $\gamma_{\rm soft,R}$.
Note that the soft non-relativistic contribution takes a simple form
\begin{align}
    \gamma_{\rm soft,NR} &=
    \frac{9}{32\pi^2}\frac{H_\text{e}^4}{m_{h,{\rm eff}}}\left(
        \frac{a}{a_\text{e}}
    \right)^{-9}
    \left[
        \sinh^{-1}\!\left(\frac{am_{h,{\rm eff}}}{k_{\rm IR}}\right)
        -\sinh^{-1}\!\left(\frac{am_{h,{\rm eff}}}{k_*}\right)
    \right],
\end{align}
where $k_{\rm IR}$ is an IR cutoff which we take $k_{\rm IR}=0.05$ Mpc$^{-1}$ in the figure.
A slight change in $k_{\rm IR}$, however, does not change the order of magnitude of the decay rate, since it  affects the result only logarithmically.

Since $\gamma_{\rm hard} + \gamma_{\rm soft,R}\propto a^{-3}$, we write $\gamma \simeq \overline\gamma \left(a/a_\text{e}\right)^{-3}$.
Then, Eq.~(\ref{eq: Boltzmann for n_h}) can be solved to find
\begin{align}
    n_h(a) &=
    n_h(a_\text{e})\left(\frac{a}{a_\text{e}}\right)^{-3}
        -\frac{2}{3}\frac{\overline\gamma\Gamma_h}{H_\text{e}}\left[
        \left(\frac{a}{a_\text{e}}\right)^{-3/2} - \left(\frac{a}{a_\text{e}}\right)^{-3}
    \right],
\end{align}
where, from Eq.~(\ref{eq: rho_h(a_e) for the stochastic fluctuation}), we have
\begin{align}
    n_h(a_\text{e}) &= \frac{3}{8\pi^2}\frac{H_\text{e}^4}{m_{h,{\rm eff}}}.
\end{align}
Supposing that $\gamma\sim\gamma_{\rm soft,R}\Rightarrow\overline\gamma\simeq\left(3/32\pi^2\right)\left(H_\text{e}^4/m_{h,{\rm eff}}\right)$ and requiring $n_h(a_{\rm dec})=0$, we find the decay time as
\begin{align}
    \frac{a_{\rm dec}}{a_\text{e}} &\simeq \left(6\frac{H_\text{e}}{\Gamma_h}\right)^{2/3}.
    \label{eq: a_dec for stochastic}
\end{align}
It is interesting that this is almost the same as the decay time of the primordial condensate given in Eq.~(\ref{eq: a_dec for misalignment}).

\section{Summary}
\label{sec: Summary}

The cosmological role of the Higgs, especially during and immediately after inflation, has two distinctive contributions, one being the primordial condensate and the other being the stochastic fluctuations.
However, the difference between the two is not well recognized and sometimes confused.
This is partly because such an elaborate separation is not necessary when one does not need to care about the momentum distribution of the primordial Higgs field.
However, if, for example, the Higgs decay is taken into account, this difference needs to be looked at more closely.

In this paper, we focused on non-thermal phase space distribution arising from these two different sources: primordial Higgs condensate and stochastic fluctuation.
Both of these components are considered as displacement of the Higgs field value during inflation and are therefore hardly distinguishable in the field space.
On the other hand, once inflation is over, these components will have different phase space distributions.

The primordial condensate can be viewed as the initial field strength of the Higgs field.
It persists during inflation and starts oscillating when the Hubble rate becomes equal to the condensate effective frequency (or mass) after the end of inflation.
The stochastic fluctuation originates from the UV modes flowing into the IR regime.
It is a secular effect if the Higgs was massless, namely, the variance of the field grows forever.
However, the existence of the potential including the self-interaction works as a drift term that tries to restore the original position of the field, namely, the origin.
The balance between these two effects leaves the finite variance for the Higgs field during inflation, by which the Higgs spectrum after inflation is affected.

The primordial Higgs condensate has the delta function-like shape of the spectrum localized at the zero momentum state.
On the other hand, the stochastic fluctuation has a nontrivial phase space distribution which consists of three different regimes depending on the comoving momentum $k$, namely, $k>m_\phi$, $k_*<k<m_\phi$, and $k<k_*$.
We derived an analytic estimate and showed that the estimate is in good agreement with our previous numerical result.

As an implication for the non-thermal Higgs spectrum, we discussed the Higgs decay after the inflation, where we assumed that the perturbative decay is always possible, while the Higgs remains non-thermal until it decays.
We found that the decay times of the primordial condensate and the stochastic fluctuation almost coincide with each other, namely, both contributions decay away almost at the same time.

Finally, we emphasize that implication of the non-thermal Higgs spectrum is not limited to Higgs decays.
Since the Higgs sector is linked to new physics in many ways, the non-thermal Higgs spectrum can have many opportunities to be relevant to Higgs cosmology.

\section*{Acknowledgements}
The authors would like to thank Sung-Mook Lee and Kenji Nishiwaki for the helpful discussion at the early stages of the project.
This work is in part supported by the JSPS KAKENHI Grant Nos.\ 19H01899 (KK and KO) and 21H01107 (KO). 
KK acknowledges support by Institut Pascal at Université Paris-Saclay during the Paris-Saclay Astroparticle Symposium 2021, with the support of the P2IO Laboratory of Excellence (program “Investissements d’avenir” ANR-11-IDEX-0003-01 Paris-Saclay and ANR-10-LABX-0038), the P2I axis of the Graduate School Physics of Université Paris-Saclay, as well as IJCLab, CEA, IPhT, APPEC, the IN2P3 master projet UCMN and EuCAPT.

\bibliographystyle{JHEP}
\bibliography{biblio}

\end{document}